\documentclass[aps,prl,amsmath,showpacs,preprintnumbers,twocolumn,amssymb,amsmath,superscriptaddress]{revtex4}

\usepackage{graphicx}
\usepackage{mathptmx, textcomp}
\usepackage[latin1]{inputenc}
\usepackage{tabularx}
\usepackage{units}
\usepackage{color}

\begin{document}
\title{Observation of Fermi surface deformation in a dipolar quantum gas}

\affiliation{Institut f\"ur Experimentalphysik and Zentrum f\"ur Quantenphysik, Universit\"at Innsbruck, Technikerstra{\ss}e 25, 6020 Innsbruck, Austria}
\author{K. Aikawa}
\affiliation{Institut f\"ur Experimentalphysik and Zentrum f\"ur Quantenphysik, Universit\"at Innsbruck, Technikerstra{\ss}e 25, 6020 Innsbruck, Austria}
\author{S. Baier}
\affiliation{Institut f\"ur Experimentalphysik and Zentrum f\"ur Quantenphysik, Universit\"at Innsbruck, Technikerstra{\ss}e 25, 6020 Innsbruck, Austria}
\author{A. Frisch}
\affiliation{Institut f\"ur Experimentalphysik and Zentrum f\"ur Quantenphysik, Universit\"at Innsbruck, Technikerstra{\ss}e 25, 6020 Innsbruck, Austria}
\author{M. Mark}
\affiliation{Institut f\"ur Experimentalphysik and Zentrum f\"ur Quantenphysik, Universit\"at Innsbruck, Technikerstra{\ss}e 25, 6020 Innsbruck, Austria}
\author{C. Ravensbergen}
\affiliation{Institut f\"ur Experimentalphysik and Zentrum f\"ur Quantenphysik, Universit\"at Innsbruck, Technikerstra{\ss}e 25, 6020 Innsbruck, Austria}
\affiliation{Institut f\"ur Quantenoptik und Quanteninformation,\"Osterreichische Akademie der Wissenschaften, 6020 Innsbruck, Austria}
\author{F. Ferlaino}
\affiliation{Institut f\"ur Experimentalphysik and Zentrum f\"ur Quantenphysik, Universit\"at Innsbruck, Technikerstra{\ss}e 25, 6020 Innsbruck, Austria}
\affiliation{Institut f\"ur Quantenoptik und Quanteninformation,\"Osterreichische Akademie der Wissenschaften, 6020 Innsbruck, Austria}

\date{\today}


\begin{abstract}
The deformation of a Fermi surface is a fundamental phenomenon leading to a plethora of exotic quantum phases. Understanding these phases, which play crucial roles in a wealth of systems, is a major challenge in atomic and condensed-matter physics. Here, we report on the observation of a Fermi surface deformation in a degenerate dipolar Fermi gas of erbium atoms. The deformation is caused by the interplay between strong magnetic dipole-dipole interaction and the Pauli exclusion principle. We demonstrate the many-body nature of the effect and its tunability with the Fermi energy. Our observation provides basis for future studies on anisotropic many-body phenomena in normal and superfluid phases.

\end{abstract}

\maketitle

\begin{figure}[t]
\includegraphics[width=1\columnwidth] {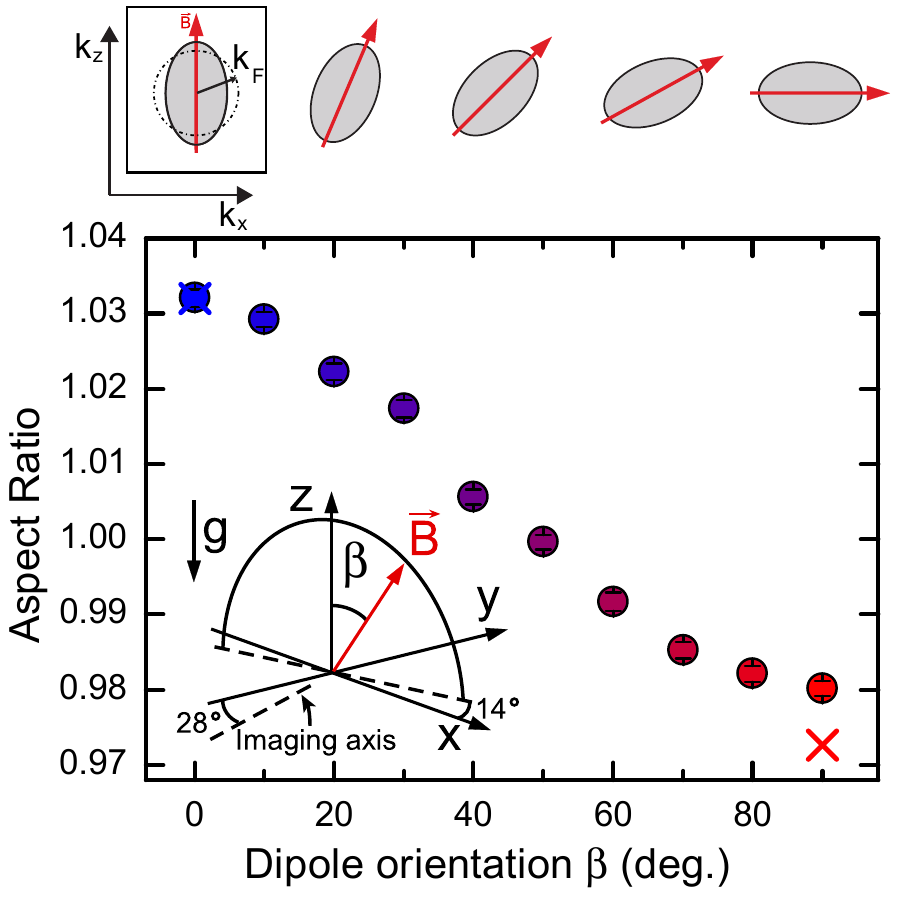}
\caption{(color online) AR of an expanding dipolar Fermi gas as a function of the angle $\beta$. In this measurement, the trap frequencies are $(f_x, f_y, f_z) = \unit[(579, 91, 611)]{Hz}$. The data are taken at $t_{\rm TOF}=\unit[12]{ms}$. Each individual point is obtained from about 39 independent measurements. The error bars indicate the standard errors of the mean. For comparison, the calculated values are also shown for $0^\circ$ and $90^\circ$ (crosses). The inset schematically illustrates the geometry of the system.  Gravity is along the $z$ direction. The atomic cloud is imaged with an angle of $28^\circ$ with respect to the $y$ axis (Supplementary Materials). The magnetic field orientation is rotated on the plane with an angle of $14^\circ$ with respect to the $xz$ plane. Schematic illustrations of the deformed FS are also shown above the panel. Here, the Fermi momentum for an ideal Fermi gas is shown as $k_F$.}
\label{fig:fig_geometry}
\end{figure}

The Fermi-liquid theory, formulated by Landau in the late 50's, is one of the most powerful tools in modern condensed-matter physics\,\cite{landau1981statistical2}. It captures the behavior of interacting Fermi systems in the normal phase, such as electrons in metals and liquid $^3$He\,\cite{leggett1975theoretical}. Within this theory the interaction is accounted by dressing the fermions as {\em quasi-particles} with an effective mass and an effective interaction. The ground state is the so-called {\em Fermi sea}, in which the quasi-particles fill one-by-one all the states up to the Fermi momentum, $k_{\rm F}$. The Fermi surface (FS), which separates occupied from empty states in $k$-space, is a sphere of radius $k_{\rm F}$ for isotropically interacting fermions in uniform space. The FS is crucial for understanding system excitations and Cooper pairing in superconductors. When complex interactions act, the FS can get modified. For instance, strongly-correlated electron systems violates the Fermi-liquid picture, giving rise to a deformed FS, which spontaneously breaks the rotational invariance of the system\,\cite{fradkin2010nematic}. Symmetry-breaking FSs have been studied in connection with electronic liquid crystal phases\,\cite{kivelson1998electronic} and Pomeranchuk instability\,\cite{pomeranchuk1959onthestability} in solid state systems. Particularly relevant is the nematic phase, in which anisotropic behaviors spontaneously emerge and the system acquires an orientational order, while preserving its translational invariance. This exotic phase has recently been observed by transport and thermodynamics studies in ruthenates\,\cite{borzi2007formation}, in high-transition-temperature superconductors such as cuprates\,\cite{ando2002electrical}, and in other systems\,\cite{fradkin2010nematic}.

A completely distinct approach to study FSs is provided by ultracold quantum gases. These systems are naturally free from impurities and crystal structures, realizing a situation close to the ideal uniform case. Here, the shape of the FS can directly reveal the fundamental interactions among particles. Studies of FSs in strongly interacting Fermi gases have been crucial in understanding the BEC-to-BCS crossover, where the isotropic $s$-wave (contact) interaction causes a broadening of the always-spherical FS\,\cite{giorgini2008theory}. Recently, Fermi gases with anisotropic interactions have attracted remarkable attention in the context of $p$-wave superfluidity\,\cite{cheng2005anisotropic,gurarie2005quantum} and dipolar physics\,\cite{baranov2012condensed}. Many theoretical studies have focused on dipolar Fermi gases, predicting the existence of a deformed FS\,\cite{miyakawa2008phase-space,fregoso2009ferronematic,fregoso2009biaxial,sogo2009dynamical,baillie2012magnetostriction,wachtler2013low}. These studies also include an extension of the Landau Fermi-liquid theory to the case of anisotropic interactions\,\cite{chan2010anisotropic}. Despite recent experimental advances in polar molecules and magnetic atoms\,\cite{griesmaier2005bose-einstein,ni2008high,lu2012quantum,aikawa2014reaching}, the observation of anisotropic FSs has so far been elusive. 

In this letter, we present the direct observation of the deformed FS in  dipolar Fermi gases of strongly magnetic erbium (Er) atoms. By virtue of the anisotropic dipole-dipole interaction (DDI) among the particles, the FS is predicted to be deformed into an ellipsoid, reflecting the underlying symmetry of the interaction for polarized gases. To minimize the system's energy, the FS elongates along the direction of the maximum attraction of the DDI, where the atomic dipoles have a 'head-to-tail' orientation. To understand the origin of the Fermi surface deformation (FSD), one has to account for both the mechanical action of the DDI in $k$-space and the Pauli exclusion principle, which imposes the many-body wave-function to be anti-symmetric. In the Hartree-Fock formalism, the FSD comes from the exchange interaction among fermions, known as the Fock term (\,\cite{miyakawa2008phase-space,baillie2012magnetostriction} and Supplementary Materials). Our observations agree very well with parameter-free calculations based on the Hartree-Fock theory\,\cite{miyakawa2008phase-space,sogo2009dynamical,wachtler2013low}. We demonstrate that the degree of deformation, related to the {\em nematic susceptibility} in the liquid-crystal vocabulary, can be controlled by varying the Fermi energy of the system and vanishes at high temperatures.

\begin{figure}[t]
\includegraphics[width=1\columnwidth] {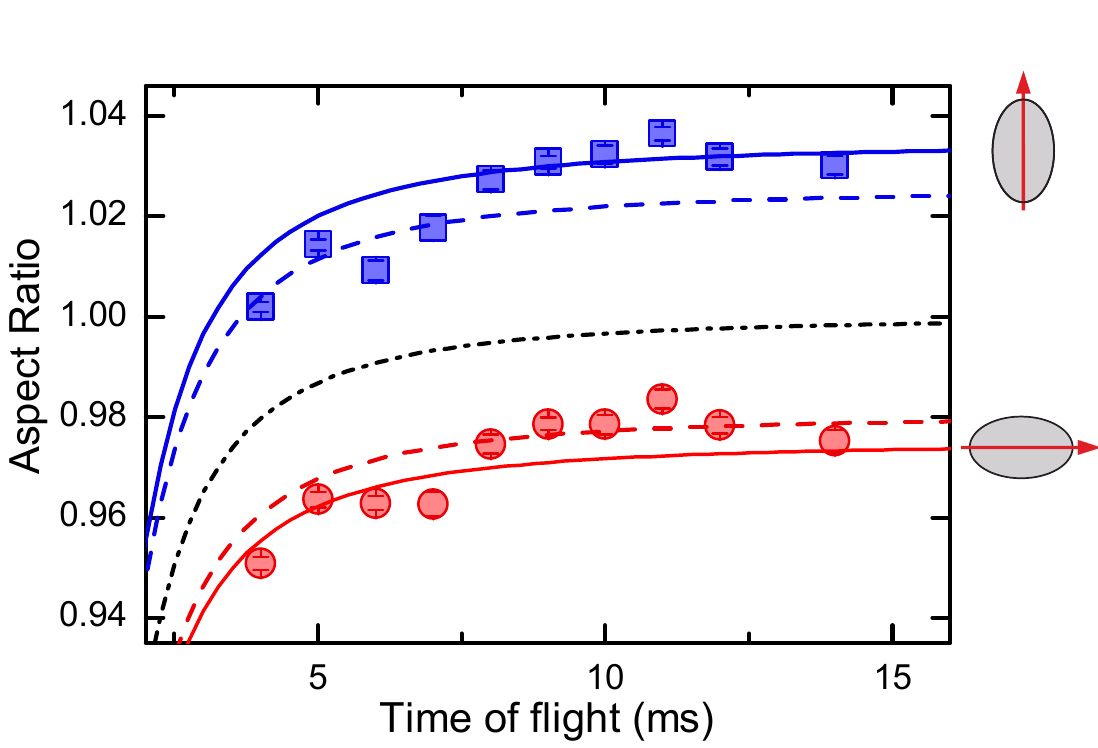}
\caption{(color online) Time evolution of the AR of the atomic cloud during the expansion. Measurements are performed for two dipole angles, $\beta = 0^{\circ}$ (squares) and $\beta=90^{\circ}$ (circles) under the same conditions as in Fig.\,1. The error bars are standard errors of the mean of about 17 independent measurements. The possible origin of the fluctuations in the AR is carefully discussed in the Supplementary Materials. The theoretical curves show the full numerical calculations (solid lines), which include both the FSD and the NBE effects, and the calculation in the case of ballistic expansions (dashed lines), i.\,e.\,in the absence of the NBE effect. For comparison, the calculation for a non-interacting Fermi gas is also shown (dot-dashed line).}
\label{fig:fig_angle_tof}
\end{figure}

Our system is a single-component quantum degenerate dipolar Fermi gas of Er atoms. Like other lanthanoids, a distinct feature of Er is its large permanent magnetic dipole moment $\mu$ (7 Bohr magneton), entailing the strong DDI among the fermions. 
Similarly to our previous work\,\cite{aikawa2014reaching}, we take advantage of elastic dipole-dipole collisions to drive efficient evaporative cooling in spin-polarized fermions. The sample is confined into a three-dimensional optical harmonic trap and typically contains $7\times 10^4$ atoms at a temperature of 0.18(1) $T_F$ with $T_F=\unit[1.12(4)]{\mu K}$ (Supplementary Materials). We control the alignment of the magnetic dipole moments  by setting the orientation of an external polarizing magnetic field. The quantity $\beta$ symbolizes the angle between the magnetic field and the $z$ axis (inset Fig.\,1).

To explore the impact of the DDI on the momentum distribution, we perform time-of-flight (TOF) experiments. Since its first use as "smoking-gun" evidence for Bose-Einstein condensation\,\cite{anderson1995observation,davis1995bose}, this technique has proved its power in revealing many-body quantum phenomena in momentum space\,\cite{bloch2008many-body,giorgini2008theory}. TOF experiments are based on the study of the expansion dynamics  of the gas when released from a trap. For sufficiently long expansion time, the size of the atomic cloud is dominated by the velocity dispersion and, in the case of ballistic (free) expansions, the TOF images purely reflect the momentum distribution in the trap.

\begin{figure}[t]
\includegraphics[width=1\columnwidth] {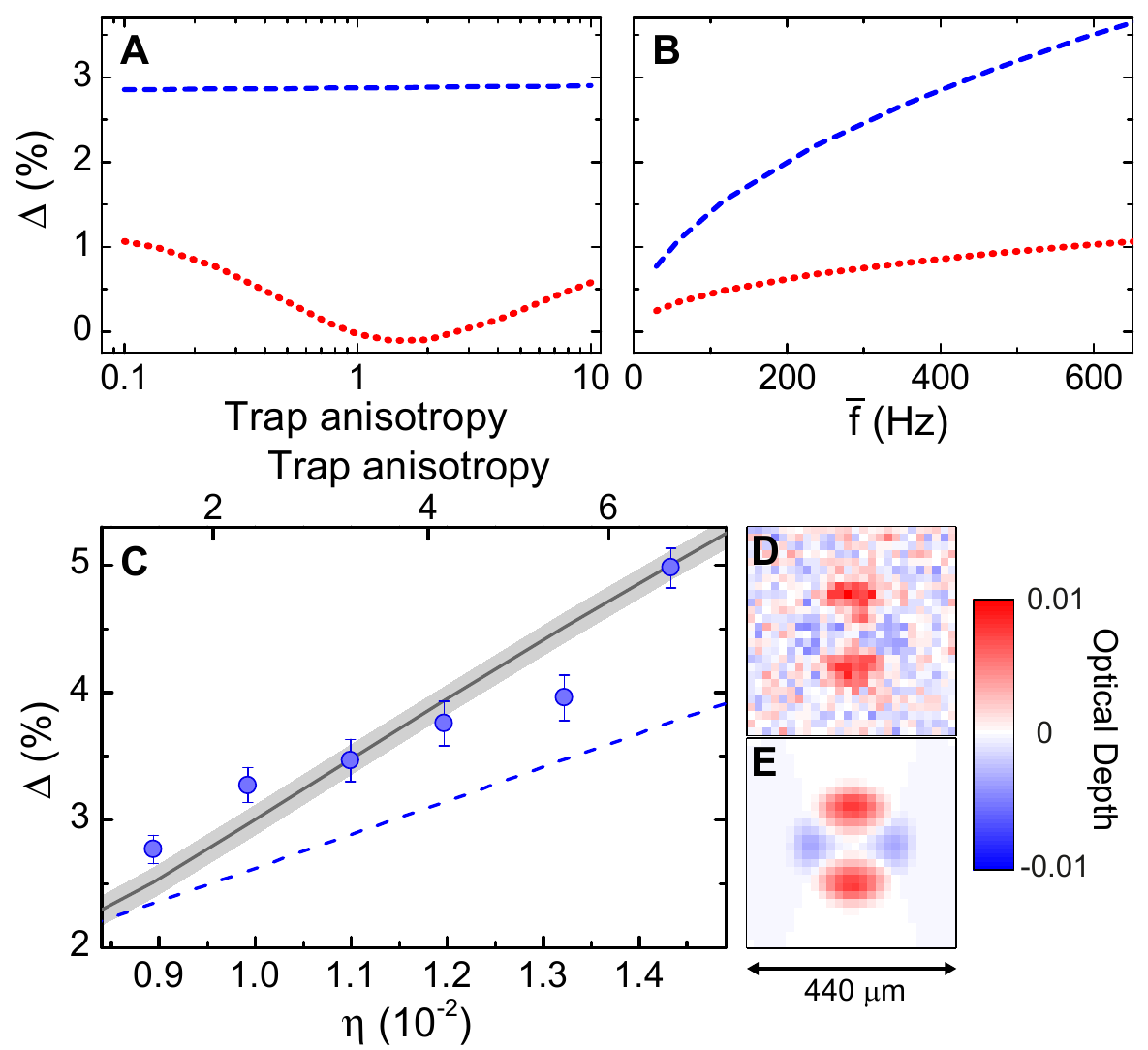}
\caption{(color online) $\Delta$ for various trap geometries. We consider a cigar-shaped trap with $f_x=f_z$ in the calculations and show the behaviors of the FSD (dashed lines) and the NBE (dotted lines) separately as a function of the trap anisotropy $\sqrt{f_x f_y}/f_z$ at $\bar{f}=\unit[400]{Hz}$ (A) and as a function of $\bar{f}$ at $\sqrt{f_x f_y}/f_z=5$ (B).
(C) Experimentally observed $\Delta$ at $t_{\rm TOF}=\unit[12]{ms}$ are plotted as a function of $\eta$, together with the full calculation (solid line) and the calculation considering only FSD (dashed line). The shaded area shows the uncertainty originating from the uncertainty in determining $\eta$ in our experiments. The sample contains $6 \times 10^4$ atoms at a typical temperature of $T/T_F = 0.15(1)$. The error bars represent standard errors of the mean of about 15 independent measurements. The variation of the trap anisotropy in the experiment is indicated in the top axis. Visualization of the FSD at $\eta=0.009$ from the experimental TOF image (D) and from the fitted image (E).}
\label{fig:fig_powdep}
\end{figure}

In our experiment, we first prepare the ultracold Fermi gas with a given dipole orientation and then we let the sample expand by suddenly switching off the optical dipole trap (ODT). From the TOF images, we derive the cloud aspect ratio (AR), which is defined as the ratio of the vertical to horizontal radius of the cloud in the imaging plane (Supplementary Materials).  Figure\,1 shows the AR for various values of $\beta$. For vertical orientation ($\beta=0^\circ$), we observe a clear deviation of the AR from unity with a cloud anisotropy of about 3 \%. TOF images show that the  cloud has an ellipsoidal shape with elongation in the direction of the dipole orientation. When changing $\beta$, we observe that the cloud follows the rotation of the dipole orientation, keeping the major axis always parallel to the direction of the maximum attraction of the DDI. In a second set of experiments, we record the time evolution of the AR during the expansion for  $\beta=0^\circ$ and $\beta= 90^\circ$ (Fig.\,2).  For both orientations, the AR differs from unity at long expansion times. Our results are strikingly different from the ones of conventional Fermi gases with isotropic contact interactions, in which the FS is spherical ($\mathrm{AR}=1$) and the magnetic field orientation has no influence on the cloud shape.

The one-to-one mapping of the original momentum distribution in the trap and the density distribution of the cloud after long expansion time strictly holds only in the case of pure ballistic expansions. In our experiments, the DDI is acting even during the expansion and could potentially mask the observation of the FSD. We evaluate the effect of the non-ballistic expansion (NBE) by performing numerical calculations  based on the Hartree-Fock mean-field theory at zero temperature and the Boltzmann-Vlasov equation for expansion dynamics\,\cite{sogo2009dynamical,wachtler2013low} (Supplementary Materials). In Fig.\,2, the theoretical curves do not have any free parameter and are calculated both in presence (solid lines) and absence (dotted lines) of the NBE effect. We observe an excellent agreement between experiment and theory, showing that our model accurately describes the behavior of the system. In addition, the comparison between ballistic and non-ballistic expansion reveals that the latter plays a minor role in the final AR, showing that the observed anisotropy dominantly originates from the FSD.

Theoretical works have predicted that the degree of deformation depends on the Fermi energy and the dipole moment\,\cite{miyakawa2008phase-space,fregoso2009biaxial,sogo2009dynamical,chan2010anisotropic,baillie2012magnetostriction,wachtler2013low}. In the limit of weak DDI, the magnitude of the FSD in a trapped sample is expected to be linearly proportional to the ratio of the DDI to the Fermi energy, $\eta=nd^2/E_F$\,\cite{baillie2012magnetostriction}. Here, $n=4\pi(2mE_F/h^2)^{3/2}/3$ is the peak number density at zero temperature with $h$ the Planck constant, $m$ the mass,  $d^2=\mu_0\mu^2/(4\pi)$ the coupling constant for the DDI, and $\mu_0$ the magnetic constant. For a harmonically trapped ideal Fermi gas, the Fermi energy $E_F$ depends on the atom number $N$ and the mean trap frequency $\bar{f}=(f_x f_y f_z)^{1/3}$, $E_F=h \bar{f} (6N)^{1/3}$. Given that $\eta \propto \sqrt{E_F}$, the FSD can be tuned by varying $E_F$.

To test the theoretical predictions, we first numerically study the degree of cloud deformation $\Delta$, defined as $\Delta={\rm AR}-1$, as a function of the trap anisotropy, $\sqrt{f_x f_z}/f_y$, and/or $\bar{f}$. To distinguish the effect of the FSD and of the NBE, we keep the two contributions separated in the calculations (Fig.\,3A and 3B). Our results clearly convey the following information: (i) the FSD gives the major contribution to $\Delta$, (ii) the FSD is independent from the trap anisotropy, while it increases with $\bar{f}$, (iii) the NBE effect is reminiscent of the trap anisotropy and vanishes for a spherical trap\,\cite{sogo2009dynamical}.

\begin{figure}[t]
\includegraphics[width=1\columnwidth] {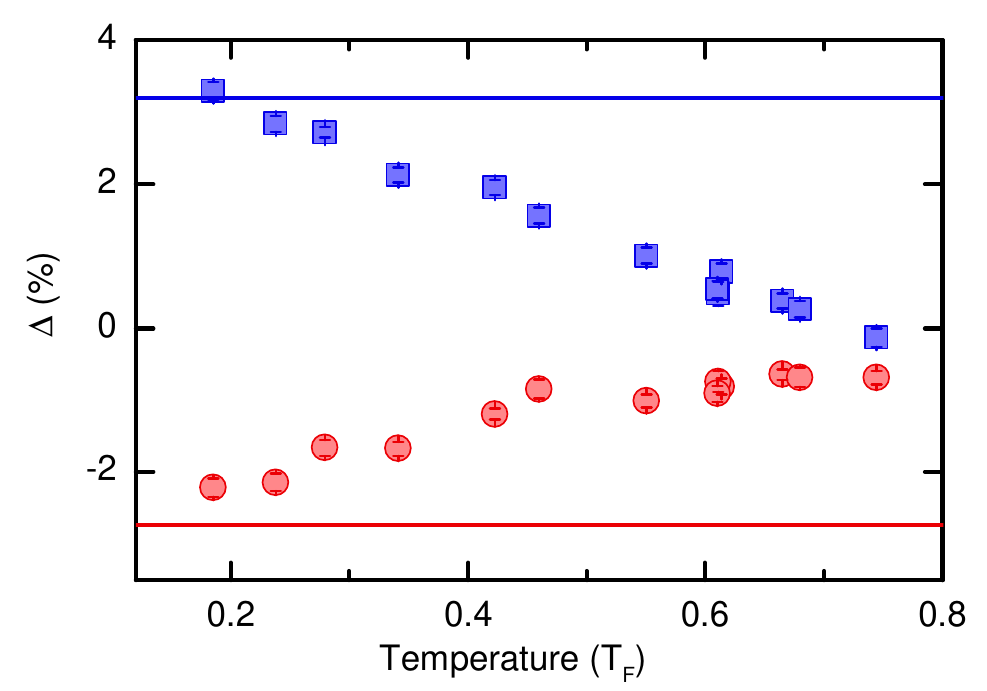}
\caption{(color online) $\Delta$ as a function of the temperature of the cloud. Measurements are performed for two dipole angles, $\beta=0^{\circ}$ (squares) and $\beta=90^{\circ}$ (circles) under the same conditions as in Fig.\,1. The error bars are standard errors of the mean of about 26 independent measurements. The solid lines show the numerically calculated values at zero temperature for $\beta=0^{\circ}$ and $\beta=90^{\circ}$. }
\label{fig:fig_tempdep}
\end{figure}

In the experiment, we explore the dependence of $\Delta$ on the trap geometry for $\beta=0^{\circ}$ by keeping the axial frequency ($f_y$) constant and varying the radial frequencies ($f_x=f_z$ within $\unit[5]{\%}$) (Fig.\,3C). This leads to a simultaneous variation of both the trap anisotropy and $\bar{f}$. We observe an increase of $\Delta$ with $\eta$, which is consistent with the theoretically predicted linear dependence\,\cite{baillie2012magnetostriction}.

In analogy with studies in superconducting materials\,\cite{rosenthal2014visualization}, we graphically emphasize the FSD in the measurements at $\eta=0.009$ by subtracting the TOF absorption image taken at $\beta=90^{\circ}$ from the one at $\beta=0^{\circ}$ (Fig.\,3D). The resulting image exhibits a clover-leaf-like pattern, showing that the momentum spread along the orientation of the dipoles is larger than in the other direction. For comparison, the same procedure is applied for images obtained by a fit to the observed cloud (Fig.\,3E). At $\eta=0.009$, the trap anisotropy is so small that the NBE effect is negligibly small and the deformation is caused almost only by the FSD.

Finally, we investigate the temperature dependence of $\Delta$ (Fig.\,4). We prepare samples at various temperatures by stopping the evaporative cooling procedure at various points. The final trap geometry is kept constant. 
When reducing the temperature, we observe the emergence of the FSD, which becomes more and more pronounced at low temperatures and eventually approaches the zero-temperature limit. 
The qualitative behavior of the observed temperature dependence is consistent with a theoretical result at finite temperatures\,\cite{baillie2012magnetostriction}, although further theoretical developments are needed for a more quantitative comparison.

Our observation clearly shows the quantum many-body nature of the FSD and sets the basis for future investigations on more complex dipolar phenomena, including collective excitations\,\cite{sogo2009dynamical,babadi2012collective,lu2013zero,wachtler2013low} and anisotropic superfluid pairing\,\cite{you1999prospects,Baranov2002spi}. Taking advantage of the wide tunability of cold atom experiments, dipolar Fermi gases are ideally clean systems for exploring exotic and topological phases in a highly controlled manner\,\cite{baranov2012condensed}.

\begin{acknowledgments}
We are grateful to A.\,Pelster, M.\,Ueda, M.\,Baranov, R.\,Grimm, T.\,Pfau, B.\,L.\,Lev, and E.\,Fradkin for fruitful discussions. This work is supported by the Austrian Ministry of Science and Research (BMWF) and the Austrian Science Fund (FWF) through a START grant under Project No. Y479-N20 and by the European Research Council under Project No. 259435. K. A. is supported within the Lise-Meitner program of the FWF.
\end{acknowledgments}

\appendix

\section{Supplementary Materials}
\subsection{Experimental setup}
We obtain a quantum gas of fermionic $^{167}$Er atoms via laser cooling in a narrow-line magneto-optical trap\,\cite{frisch2012narrow} followed by evaporative cooling in an ODT\,\cite{aikawa2014reaching}. The sample is trapped in a crossed ODT consisting of a horizontally ($y$ axis) and a vertically ($z$ axis) propagating beam at $\unit[1570]{nm}$. The beam waist of the horizontal beam is $\unit[15]{\mu m}$, while the one of the vertical beam is tuned in a range from $\unit[20]{\mu m}$ to $\unit[90]{\mu m}$ to vary the trap geometry from a nearly spherical shape to a nearly cigar shape. During the entire experimental procedure, the fermions are fully polarized into the lowest hyperfine sublevel $|F=19/2, m_F=-19/2\rangle$, where $F$ is the total angular momentum quantum number and $m_F$ is its projection along the quantization axis. For maintaining the spin polarization of the trapped sample, we apply an external magnetic field of $\unit[0.58]{G}$. At this field value, we do not observe any influence of Feshbach resonances\,\cite{frisch2014quantum}. The magnetic field orientation is controlled with two sets of coils. During evaporative cooling, the magnetic field is vertically oriented ($\beta=0^\circ$). For imaging, we rotate the magnetic field orientation to the direction of the imaging axis to attain a maximum optical depth.

\begin{figure}[t]
\includegraphics[width=1\columnwidth] {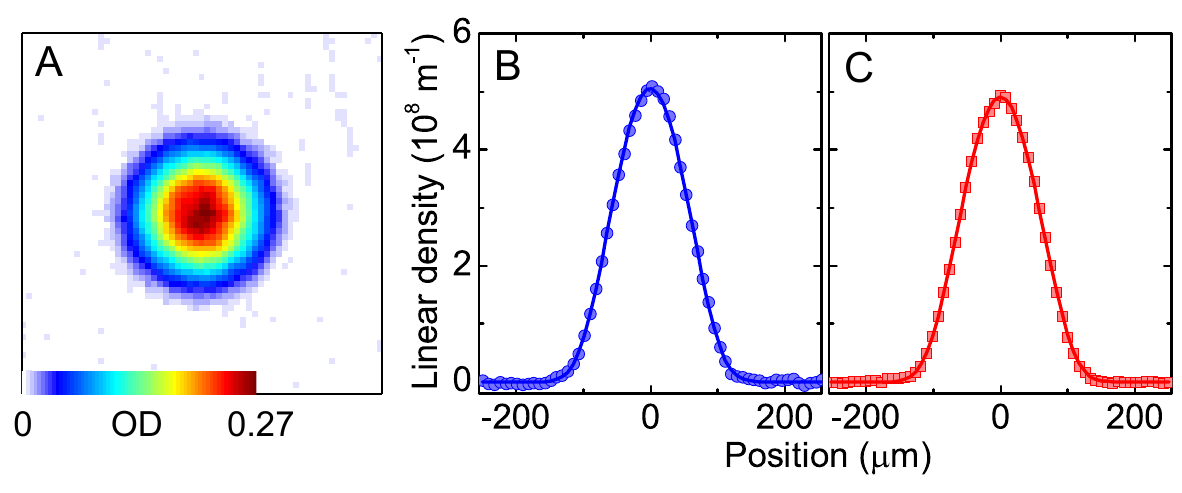}
\caption{(color online) Time-of-flight absorption image (A) and its integrated profiles in the horizontal (B) and vertical (C) directions. The image corresponds to the measurement at $\beta=0^\circ$ in Fig.\,1 and is averaged over 39 independent measurements. The integrated profiles of the fit with a poly-logarithmic function are shown by solid lines. The AR of the cloud is about 1.03.}
\label{fig:fig_tofimage}
\end{figure}

\subsection{Measurement of the AR}
We measure the deformation of the cloud shape in TOF absorption images by using a standard poly-logarithmic fit for the integrated density distribution of an ideal Fermi gas\,\cite{Inguscio2006ufg,desalvo2010degenerate,tey2010double}
\begin{equation}
n(X,Z)=B {\rm Li}_2\Biggl(-\zeta {\rm exp}\biggl(-\frac{(h-h_0)^2}{2\sigma_h^2}-\frac{(z-z_0)^2}{2\sigma_z^2}\biggr)\Biggr)
\label{eq2}
\end{equation}
where Li$_n$ is the $n$-th order poly-logarithmic function, $\zeta$ is the fugacity, $h$ and $z$ indicate horizontal and vertical coordinates on the imaging plane, respectively, $h_0$ and $z_0$ are the positions, and $\sigma_h$ and $\sigma_z$ are the radii. We define the AR of the observed cloud as $\sigma_z/\sigma_h$. The imaging axis has an angle of $28^\circ$ with respect to the $y$ axis (inset Fig.\,1) and thus the horizontal radius in the imaging plane $\sigma_h$ is related to the radii in the $x$ and $y$ directions, $\sigma _x$, $\sigma _y$, by
\begin{equation}
\sigma_h = \sqrt{\sigma_x^2 {\rm cos}^2(28^\circ) + \sigma_y^2 {\rm sin}^2(28^\circ)}
\label{eq3}
\end{equation}

The fugacity $\zeta$ is directly connected to $T/T_F$ through the relation $T/T_F=[-6\times {\rm Li}_3(- \zeta)]^{-1/3}$. The optical depth is proportional to $N$. Although the optical depth is also related to $\zeta$ through $N$ and $T_F$ by $T_F=h \bar{f} (6N)^{1/3}/k_B$, we leave both free in the fitting procedure and confirm that they are consistent with each other. Here, we assume that the cloud has a constant fugacity over the entire cloud because the momentum deformation is small. Rigorously speaking, $T_F$ is anisotropic and $T$ is constant over the cloud, and thus $\zeta$ should be anisotropic. Dealing with such a distribution is beyond the scope of the present work. Figure\,5 shows a typical TOF absorption image and its integrated profiles as well as the integrated profiles of the poly-logarithmic fit. The fit is in excellent agreement with the observed distribution.

By taking the average of about 20 independent measurements, we are able to determine the AR with a typical precision of $\unit[0.1]{\%}$, corresponding to the standard error of the mean for multiple measurements. In addition, we find six possible sources of systematic errors in the measured AR. (a) Variation in pixel sizes. The variation in pixel sizes over the area of the cloud can introduce a systematic error in the AR. There is no measured data available for our CCD camera (Andor, iXon3). (b) Residual interference fringes. Interference fringes, arising from dusts on the imaging optics, can produce a fixed background pattern on the image. (c) Finite pixel number. The finite number of pixels can limit the resolution of the measurement of the AR, in particular at short TOF. From the TOF measurements shown in Fig.\,2, where the position of the atomic cloud varies with TOF by a free fall, we estimate the combined effect of (a), (b), and (c) to be within $\pm \unit[0.5]{\%}$. (d) Error in the fitting procedure. Although our fitting procedure assuming a constant fugacity may give rise to a systematic error in deriving the AR, it is difficult to quantitatively estimate it owing to the lack of an appropriate model. Investigating this effect will be an important future work. (e) Fluctuations in the magnetic field. The influence of the fluctuation in magnetic field, which results in a fluctuation in the dipole orientation, is negligibly small at $\beta=0^{\circ}$ and $\beta=90^{\circ}$ ($<\unit[0.05]{\%}$ in deformation). (f) Tilt of the camera. Assuming that the camera is aligned perpendicular to the imaging beam path within 1$^{\circ}$, we infer that the influence of the tilt of the camera on the AR is negligible ($<\unit[0.02]{\%}$).

\subsection{Physical origin of the deformation of the FS}
Within the Hartree-Fock theory for a many-body system, the DDI contributes to the total energy of the system in two distinct ways: the Hartree direct interaction and the Fock exchange interaction\,\cite{miyakawa2008phase-space,baillie2012magnetostriction}. As compared to the case with a non-interacting gas, the Hartree term gives rise to a distortion in position space, whereas the exchange term gives rise to a distortion in momentum space. Previously, magnetostriction in position space was observed in a dipolar BEC of chromium atoms\,\cite{stuhler2005observation}. In a BEC, the Fock term is zero because of the symmetric character of the many-body wave function. In an isotropically interacting Fermi gas, the Hartree and the Fock terms cancel out\,\cite{miyakawa2008phase-space}. The existence of the exchange term in dipolar Fermi gases arises from the combined effect of the DDI and the Pauli exclusion principle. In our expansion measurements, both the Hartree and the Fock terms need to be considered. The first is responsible for the NBE, while the second gives the FSD. 

\subsection{Calculation of the deformation}
In the present work, the collision rate associated with universal dipolar scattering\,\cite{bohn2009quasi-universal,aikawa2014reaching} is lower than the lowest trap frequency. Therefore, our sample is in the collisionless regime, where the mean free path is longer than the size of the cloud\,\cite{Pethick2002book}. 
We describe the trapped dipolar Fermi gas in the collisionless regime in the zero temperature limit with an ansatz that the Wigner distribution function is given as an ellipsoid 
\begin{equation}
g({\bf r},{\bf k},t)=\Theta \Biggl(1-\sum^{3}_{j=1} \frac{r_j^2}{R_j^2}-\sum^{3}_{j=1} \frac{k_j^2}{K_j^2}\Biggr)
\end{equation}
where $\Theta$ denotes the Heaviside's step function, and ${\bf r}$, ${\bf k}$, and $t$ denote coordinate, wave vector, and time, respectively. The parameters $R_j$ and $K_j$ represent the Thomas-Fermi radius and the Fermi momentum in the $j$th direction, respectively. These parameters are numerically determined by minimizing the total energy in the presence of the DDI. The validity of this approach was numerically confirmed\,\cite{ronen2010zero}. At equilibrium, the parameters $K_j$ contain the information of the anisotropic FS.

The expansion dynamics is calculated using the Botzmann-Vlasov equation for the Wigner distribution function under the scaling ansatz\,\cite{guery2002mean,menotti2002expansion,hu2003expansion}. The scaling parameters, representing variations from the equilibrium condition, are described by a set of coupled time-dependent differential equations. The NBE effect is naturally included in this framework and occurs predominantly within $\unit[1]{ms}$ after the release from the trap. We numerically solve the equations for the general triaxial geometry, where the trap frequencies in three directions are different and the dipoles are oriented in the direction of one of the trap axes. This reflects our experimental situation at $\beta=0^{\circ}$. Although at $\beta=90^{\circ}$ the dipole orientation has an angle of $14^\circ$ with respect to the $x$ axis, we assume that the dipole orientation is parallel to the $x$ axis in our calculation. 

We calculate the radii of the cloud on the image plane, taking into account the angle of $28^\circ$ between the imaging axis and the $y$ axis by using eq.\,(2).  In all our measurements, we observe an asymmetry between $\beta=0^{\circ}$ and $\beta=90^{\circ}$, i.\,e.\,$|\Delta|$ is larger at $\beta=0^{\circ}$ than at  $\beta=90^{\circ}$. We observe this asymmetry also in the subtracted images in Fig.\,3D and Fig.\,3E as a higher contrast in the vertical direction than in the horizontal direction. This asymmetry is well reproduced by our calculation and is understood as follows. At $\beta=0^{\circ}$, the major axis of the ellipsoid is oriented to the $z$ direction and is fully imaged. By contrast, at $\beta=90^{\circ}$, the major axis is not perpendicular to the imaging plane and we observe a combined radius between the major and the minor axis of the ellipsoid. Therefore, the observed deformation at $\beta=90^{\circ}$ is always smaller than the one at $\beta=0^{\circ}$.

\subsection{Image subtraction for Fig.\,3D,E}
The image shown in Fig.\,3D is obtained as follows. The TOF absorption images from 18 independent measurements are averaged and binned by $2\times2$ pixels to reduce background noise. This procedure is applied for the measurements at $\beta=0^{\circ}$ and $\beta=90^{\circ}$. We subtract the image at $\beta=90^{\circ}$ from the one at $\beta=0^{\circ}$. This image subtraction is very sensitive to the relative position of the clouds on the two images down to a sub-pixel level. We obtain accurate positions of the center of the cloud from the fit and shift the coordinate of the image at $\beta=90^{\circ}$ such that the center positions of two images exactly agree. We then apply spline interpolation for the image at $\beta=90^{\circ}$ to estimate the optical depth of the cloud at each pixel position in the image at $\beta=0^{\circ}$. Unlike the procedure used in Ref.\,\cite{rosenthal2014visualization}, where the anisotropy is extracted by rotating a single image by 90$^{\circ}$ and subtracting it from the original image, our procedure with two images at two dipole orientations allows us to extract only the anisotropy originating from the DDI.

\bibliographystyle{apsrev}

\bibliography{Er_FSD_ref,ultracold}

\end{document}